\documentclass[a4paper,onescolumn,twoside,fleqn,10pt]{article}
\usepackage{amsmath}
\usepackage{authblk}
\usepackage[svgnames]{xcolor}
\definecolor{blue}{RGB}{0,0,255}
\usepackage{caption}
\usepackage[font={color=blue},figurename=Fig.,labelfont={it}]{caption}
\usepackage{bm}
\usepackage{subcaption}
\usepackage[varg]{txfonts}
\usepackage{graphicx}
\usepackage{pdfpages}
\usepackage{wasysym}
\usepackage{rotating}
\usepackage{hyperref}
\usepackage{indentfirst}
\usepackage{float}
\usepackage{lipsum} 
\usepackage{setspace} 
\usepackage{indentfirst}
\usepackage[english]{babel}
\usepackage[switch,columnwise]{lineno}
\usepackage[left=2cm,right=2cm,top=2cm,bottom=2cm]{geometry}
\usepackage{cite}
\usepackage{enumitem}   

\hypersetup{
    colorlinks=true,                          
    linkcolor=blue, 
    citecolor=blue, 
    urlcolor=blue,
    linktoc=all
}

\newcommand{\blue}[1]{\textcolor{blue}{#1}}

\newcommand{\eg}{\textit{eg. }}
\newcommand{\cf}{\textit{cf. }}

\newcommand{\degree}{$^{\circ}$}

\newcommand{\Lamb}{\blue{Lamb} }
\newcommand{\Lopes}{\blue{Lopes et al.} }
\newcommand{\BEL}{\blue{Brückner-Egeson-Lockyer} }
\newcommand{\Milankovic}{\blue{Milankovi\'c} }
\newcommand{\Arrhenius}{\blue{Arrhenius} }
\newcommand{\Schwabe}{\blue{Schwabe} }
\newcommand{\Pouillet}{\blue{Pouillet} }
\newcommand{\Laplace}{\blue{Laplace} }
\newcommand{\Lagrange}{\blue{Lagrange} }
\newcommand{\Taylor}{\blue{Taylor-Couette} }
\newcommand{\Allan}{\blue{Allan and Ansell } }
\newcommand{\Hadley}{\blue{Hadley} }
\newcommand{\Ferrel}{\blue{Ferrel} }
\newcommand{\Douglas}{\blue{Douglas} }
\newcommand{\Rayner}{\blue{Rayner et Osborn} }
\newcommand{\Fourier}{\blue{Fourier} }
\newcommand{\Keogh}{\blue{Keogh} }
\newcommand{\Marchi}{\blue{De Marchi} }

\usepackage{fancyhdr}
\begin{document}
\title{On the tilt of the Earth's polar axis ($\kappa\lambda\iota\mu\alpha$): Some 'impressionist' remarks}

\author[1]{Courtillot Vincent}
\author[1]{Lopes Fernando}
\author[4]{Kossobokov Vladimir} 
\author[5]{Zuddas Pierpaolo}
\author[2]{Gibert Dominique}
\author[3]{Boul\'e Jean-Baptiste}
\author[1]{Le Mouël Jean-Louis}

\affil[1]{Universit\'e Paris Cité, Institut de Physique du globe de Paris, CNRS UMR 7154, Paris, France}
\affil[2]{LGL-TPE - Laboratoire de Géologie de Lyon - Terre, Planètes, Environnement, Lyon, France}
\affil[3]{CNRS UMR7196, INSERM U1154, Museum National d'Histoire Naturelle, Paris, France}
\affil[4]{Institute of Earthquake Prediction Theory and Mathematical Geophysics, Moscow, Russia}
\affil[5]{Sorbonne Universit\'e, METIS, UMR 7619, 4 Place Jussieu, 75005, Paris, France}

\date{\today}

\maketitle
	\abstract{In this lengthy letter, we wanted to discuss the concept of climate based on definitions established for over a century and direct observations that we have been collecting for more than a century as well. To do this, we present and discuss the remarkably stable maps over time of the various physical parameters that make up the climate corpus: solar temperature, atmospheric pressure, winds, precipitation, temperature anomalies. This impressionistic tableau that we are gradually sketching as our reflection unfolds leads us to the following proposition: What if, as Laplace first proposed in 1799 and later Milankovic in 1920, ground temperature were merely a consequence of climate and not a separate parameter of climate in its own right?}

\section{\label{sec01} Introduction}
	In ancient Greek, $\kappa\lambda\iota\mu\alpha$ referred to the inclination of the ground and, on a larger scale, to the inclination of the Earth towards the pole starting from the equator, hence the concept of climate (\cite{bailly1950}, p.1102). It was well understood that there was a connection between the angle at which the sun's rays intersect a given point on the Earth's surface (relative to the local vertical) and the climate of the region in question. In his work, which remains a reference to this day, \Lamb (\cite{lamb2013}, page 5) defines climate as the average of meteorological phenomena, taken over several decades. This average is characterized by a trend, a standard deviation from this trend, as well as extreme phenomena associated with the trend. \Lamb (\cite{lamb2013}, page 4) lists these meteorological phenomena: they include temperature, atmospheric pressure, humidity, clouds, precipitation, sunshine, and this list is not exhaustive.
	
	Each one of these phenomena can be considered individually. The question that arises then is to determine the 'appropriate' duration for this time average in order to distinguish as efficiently as possible the domain of climate from that of meteorology. A 'universal' agreement was reached during the 1935-Warsaw and 1957-Washington international meteorological conferences (\cf \cite{monin2000}). A duration of 30 years was chosen, based on the observation of a 'natural' cycle lasting 30 years in atmospheric circulation (\eg \cite{raspopov2000,tomasino2000,klyashtorin2007,courtillot2013}), known as the \BEL cycle (BEL cycle, \cf \cite{bruckner1890}).
	
	In our lengthy letter, we will systematically address each of the physics that define the climate: solar temperature (Section \ref{sec02}), variations in atmospheric pressure (Section \ref{sec03}) linked to the Earth's axis rotation. Next, we will discuss gradient maps of this pressure (Section \ref{sec04}), which give rise to both the circulation cells of atmospheric air masses (Hadley, Ferrel, polar, see Section \ref{sec05}) and are, with the ground topography, associated  with climate indices (Section \ref{sec06}). We will then treat about precipitation maps (Section \ref{sec07}) and conclude with arguably the most debated parameter, the temperature anomaly maps (Sections \ref{sec08} to \ref{sec11}). We will conclude our letter with an in-depth discussion and a reevaluation of the theories proposed by \Milankovic (1920, \cite{milankovic1920}) and \Arrhenius (1896, \cite{arrhenius1896}).

\section{\label{sec02} On temperature and the Sun}
	One of the parameters mentioned by \Lamb (\cite{lamb2013}) is temperature, that one might first assume to have its origin in the flow of energy and matter from the Sun. However, there is no obvious 30-year cycle in temperature, or to be more precise, this cycle remains of relatively low amplitude (\cf \cite{lovejoy2013,lemouel2020a}). This observation can be extended to solar cycles, such as the sunspot cycles (\eg \cite{usoskin2017,courtillot2021}), that represent part of the Sun's activity. Sunspots, with observations that date back to the early 18th century, exhibit time variations that, from a spectral analysis perspective, can be decomposed into several quasi-periodic cycles, among which the 11-year component, known as the \Schwabe (\cite{schwabe1844}) cycle, is the best known. Yet, the signature of this cycle is only weakly present in global average temperature records (\eg \cite{kossobokov2010,lemouel2010,kossobokov2012,kossobokov2019,scafetta2016,nogueira2019,lemouel2020a,courtillot2023a}), whereas it is clearly (if paradoxically) visible, for example, in dendrochronological analyses (\eg \cite{sinclair1993,rigozo2002,brehm2021,courtillot2023b}), that are widely used in the study of past climates (\eg \cite{martinelli2004,sheppard2010,fritts2012,correa2019}).

	One might ask whether the significance of temperature in what is consider to be climate should be reconsidered. Isn't discussing variations in the temperature of the Earth’s lower atmosphere (which all agree have a significant portion originating from astronomy and astrophysics) while also mentioning a 'solar constant'\footnote{The flow of energy from the Sun at the distance of Earth's orbit} a contradiction? The solar "constant" was first estimated by \Pouillet (\cite{pouillet1837}) to be 1230 W.m$^{-2}$. It is now estimated to average about 1366 W.m$^{-2}$, with a cyclic variation of approximately 0.1 to 0.3\% about its mean over an 11-year period (\eg \cite{eddy1982,kuhn1988,li2012}). 
	
		In his Treatise on Celestial Mechanics, \Laplace (\cite{laplace1799}, vol. 5, chap. 1, page 347) wrote:
“\textit{We have shown ($\ldots$) that the mean rotational motion of the Earth is uniform, assuming that this planet is entirely solid, and it has just been demonstrated that the fluidity of the sea and the atmosphere should not alter this result. The motions induced by the Sun's heat in the atmosphere, giving rise to the trade winds, would seem to diminish the Earth's rotation: these winds blow between the tropics from west to east, and their continuous action on the sea, the continents, and the mountains they encounter appears to gradually weaken this rotational movement. However, the principle of conservation of areas shows us that the overall effect of the atmosphere on this motion must be negligible; for the solar heat, by equally expanding the air in all directions, should not alter the sum of the areas described by the vector radii of each molecule of the Earth and the atmosphere, multiplied respectively by their corresponding molecules. This requires that the rotational motion not be diminished. Thus, we are assured that while the analyzed winds reduce this motion, the other atmospheric movements that occur beyond the tropics accelerate it by the same amount}”.

	In this passage, the French scientist clearly considers the effect of variations in solar temperature, and even temperature in general, on the movements of the Earth's rotational pole to be negligible. This is in line with \Milankovic (\cite{milankovic1920})'s theory and holds true regardless of the duration of the observation period (\eg \cite{lopes2022a}). \Laplace (\cite{laplace1799}) and \Lagrange (\cite{lagrange1788}) were convinced that the movements of the mobile parts of our planet stem from the overall motion of the Earth, which is influenced by astronomical phenomena. Furthermore, they recognized the direct action of astronomical forces on the oceans, with coastal tides being the primary example.

\section{\label{sec03} On Rotation, Pressure, Flow and Winds}
	The law of conservation of areas mentioned by \Laplace (\cite{laplace1799}) can be interpreted as representing the balanced exchange of angular momentum through friction at the boundary between the atmosphere and the continents. In this specific case, the system of the ‘solid Earth and atmosphere’ can be considered as a system of concentric rotating spheres. Indeed, according to \Laplace (\cite{laplace1799}), the effect of seas and oceans on the Earth's overall motion is equivalent to the effect these same oceans would have if they were solid. The flow resulting from such a stationary configuration is known as \Taylor flow (\textit{cf} \cite{taylor1923}). While a formal solution has long been available for concentric cylinders (\cf \cite{landau1987}, Chapter 27, pages 99 to 103), obtaining a formal mathematical solution for concentric spheres remains challenging even today (\eg \cite{mannix2021}). The steady-state of this flow is characterized by symmetric patterns about the axis of rotation (both for spheres or cylinders) known as the Taylor rolls.
	
	We recently had the opportunity to test this last assertion of \Laplace (\cite{laplace1799}) by examining in detail the space-time evolution of maps of the trend of sea-level atmospheric pressure (SLP) since 1850 (\cf \cite{lopes2022b}), as well as the space-time evolution of maps of seasonal oscillations extracted from SLP, also since 1850 (\cf \cite{courtillot2022}). In the first study, we observed the presence of a time-stable Taylor-Couette pattern (consistent with flow solutions) around the rotational poles in both the Northern and Southern Hemispheres, which we named 'Triskeles'. As for the second study, it allowed us to highlight symmetry rolls around the axes of rotation of the poles and also revealed the existence of a harmonic series of seasonal forcing in the observations (1 year, 1/2 year, 1/3 year, 1/4 year, and 1/5 year), which is another manifestation of a \Taylor flow solution.
	
\begin{figure}[H]
    \centering
    \includegraphics[width=1\columnwidth]{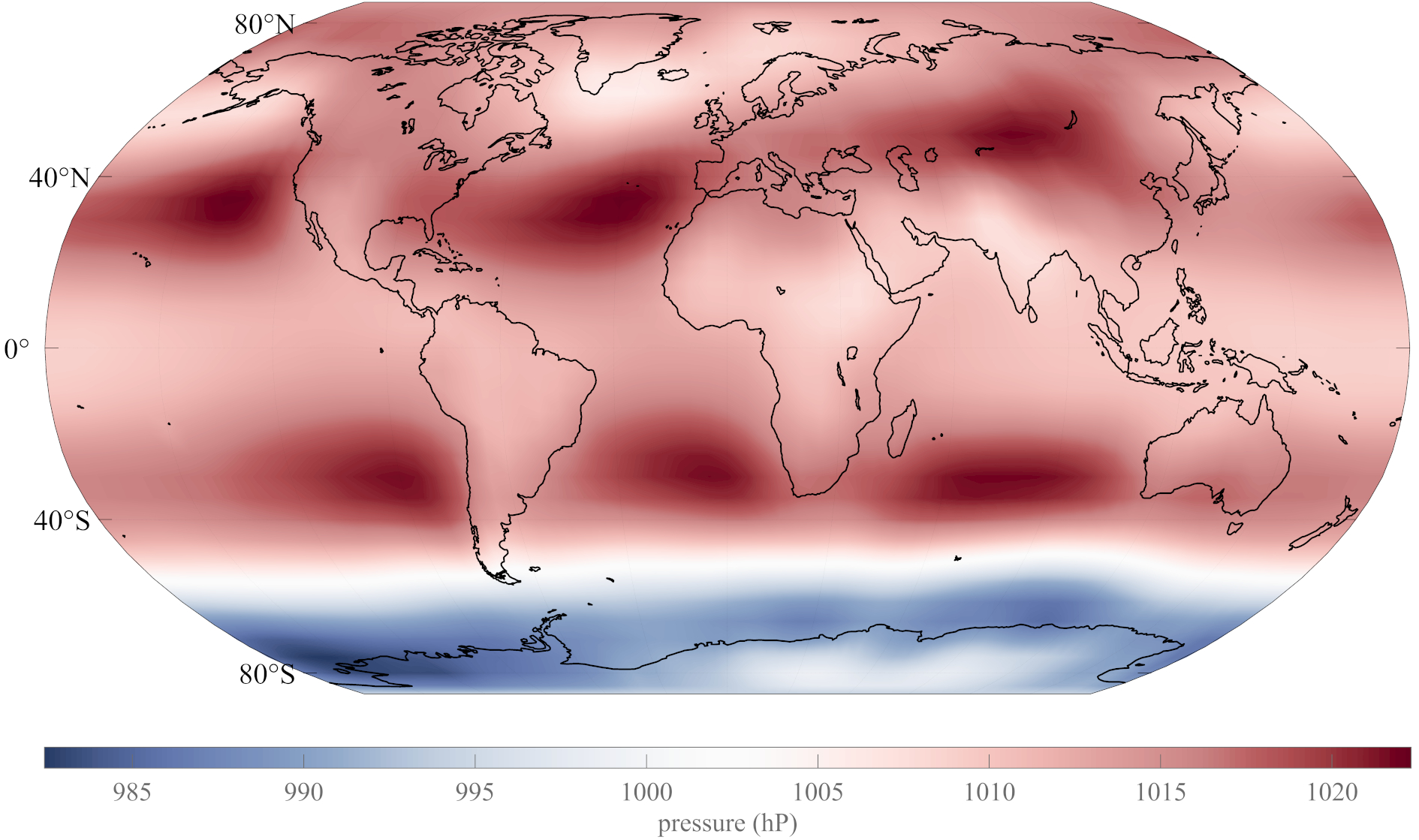}	
    \caption{Map of mean sea-level pressure trends since 1850. Robinson projection.}
	\label{fig:01}
\end{figure}

	These patterns are often considered to correspond to the existence of winds in the locations associated with these patterns. In order to assess this, we started with raw observations, namely, monthly maps of sea-level atmospheric pressure (SLP) since 1850, with a resolution of 5\degree by 5\degree. These maps are provided by the Met Office Center1, and their production is described by \Allan (\cite{allan2006}). In Figure \ref{fig:01}, we present the mean value extracted from these SLP maps: this mean value has fluctuated by less than 0.1\% since 1850.
\newpage

\section{\label{sec04} The Pressure Gradient}
\begin{figure}[H]
    \centering
    \includegraphics[width=1\columnwidth]{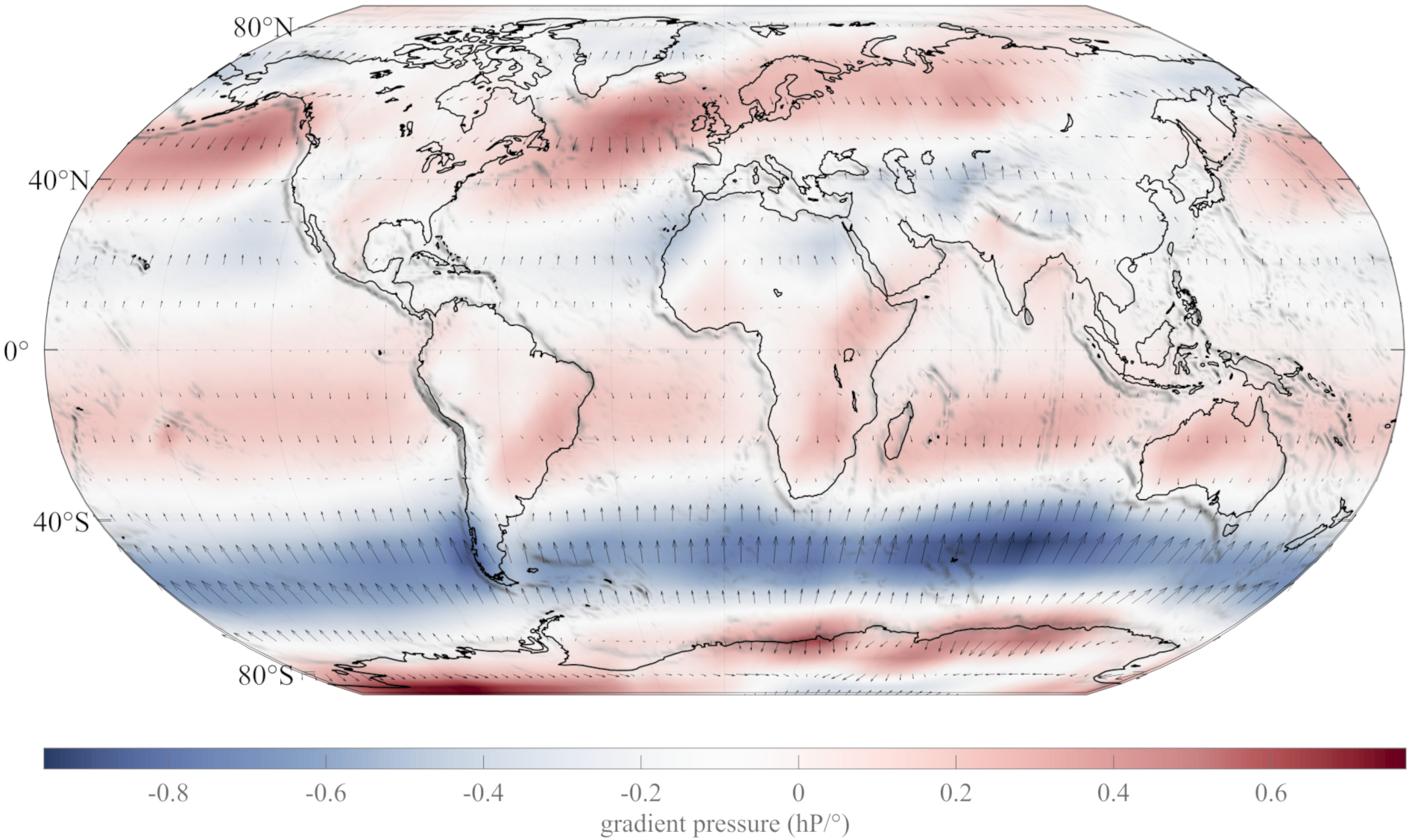}	
    \caption{Map of gradients of mean sea level pressure superimposed on the map of gradients of topography (black). Robinson projection.}
	\label{fig:02}
\end{figure}

	In Figure \ref{fig:02}, we have calculated the gradient of the average sea-level pressures and superimposed it on the map of the gradient of topography in order to determine the location of the friction zones mentioned earlier.
	
	As expected, there is a strong relationship between the pressure gradients and the orientation and magnitude of the man mountain ranges. The pressure gradients organize themselves into large structures, often oriented east-west and bounded by topographic gradients, which are often oriented north-south. Moving from west to east, one can clearly observe how pressure gradient anomalies, both negative (blue) and positive (red), encounter the American Cordillera. From north to south, this occurs for the Brooks and Alaska ranges, then in Canada along the Yukon, and finally in America along the Rockies. Similar arrangements are found along the Andes mountain range, from Guatemala to southern Patagonia.
\newpage

	Another example, which partly explains the mystery of the Asian monsoon (\eg \cite{tardif2023}), extends from India to Mongolia. Although weak in the Indian Ocean, the pressure gradient anomaly in this region is positive and changes sign after crossing the Himalayas. The boundary is particularly well marked by the topography gradient. This significant role of topography is of course well-known (\eg \cite{lamb2013}, pages 84 to 87).

\section{\label{sec05} Atmospheric Cells and Climate Change}
	The pressure gradient forms zonal bands, alternating in sign approximately every 30\degree of latitude, similar to the commonly accepted distribution of atmospheric circulation cells: the \Hadley (\cite{hadley1735}) cell between the equator and $\pm$30\degree, the \Ferrel (\cite{ferrel1856}) cell between latitudes $\pm$30\degree and $\pm$60\degree, and the polar circulation cells beyond $\pm$60\degree. The mechanism generally invoked to explain these cells is based on thermal exchanges (\cf \cite{lamb2013}, Chapter 3). At the equator, where the temperature is highest, air masses expand and rise to the tropopause, at about 15 km altitude, where temperature reaches a minimum. High-altitude air masses are pushed towards the tropics, gradually cool, and initiate their descent about $\pm$30\degree. At polar latitudes, downward convection drives air masses toward the equator, that warm in the process. The combined action of the Hadley and polar cells gives rise to the mid-latitude \Ferrel cells.
	
	One has to take into account an observed increase in the anomaly of average surface temperature across the globe (\cf \cite{masson2021}, \eg \cite{lopes2022a,courtillot2023a}), at least since the early 1980s, primarily in the Northern Hemisphere. This increase, on the order of a degree Celsius, results in the melting of Arctic sea ice, while the area of Antarctic sea ice has been increasing (but to a lesser extent; \eg \cite{holland2012,parkinson2012,desantis2017,lemouel2021, lopes2023a}). This should lead to changes in the latitude and size (location and dimensions) of the three types of cells. Although this issue has been a subject of discussion (\eg \cite{chang1995,dima2003,frierson2007,hu2007}), the recent warming does not seem to be reflected in all observations, as evidenced by the stability of sea-level atmospheric pressure trends since 1850 (\cf Figures \ref{fig:01} and \ref{fig:02}).
	
\section{\label{sec06} Climate Indices}
	Climate indices have been developed to study the impact of oceans and the atmosphere on Earth's climate. These indices take the form of time series that describe the behavior of specific regions of the oceans and the Earth's atmosphere (\eg \cite{steinbach2003,lemouel2019a,mares2022}). Among the most commonly used indices are El Niño, a warm coastal current off the coast of Peru, the PDO (Pacific Decadal Oscillation), which describes surface temperature variations in the Pacific Ocean basin (\cf \cite{mantua1997}), and the SAM (Southern Annular Mode), which is generally defined as the difference in mean zonal sea-level pressure between latitudes 40\degree S and 65\degree S (\cf \cite{marshall2003}).
	
	All these indices, which are based on temperatures, marine pressures, rainfall, as well as the cyclicity and nature of winds recorded at sea level, can be accessed in a comprehensive list available on the NOAA website1. We seek to highlight potential interconnections between these geophysical variables, interconnections for which there may not necessarily exist a physical model. The idea of subjecting these observations to multivariate statistical analysis methods, such as PCA (Principal Component Analysis, \cf \cite{hotelling1933}) or SSA (Singular Spectrum Analysis, \cf \cite{vautard1989}), naturally comes to mind. Both of these statistical methods benefit from the power of Singular Value Decomposition (SVD, \cf \cite{golub1971}). They can be summarized as follows.
	
	Let $\mathcal{D}$ be a data matrix; its singular value decomposition is $\mathcal{D}= U\lambda V^{\star}$. The principal columns of matrix $U$ represent the trends of the entire dataset; the vectors in $U$ represent the 'directions of greatest variation' in the dataset. The diagonal values of $\lambda$ are comparable to the ‘energy’ or ‘representativeness’ of these components. SVD allows the construction of an empirical model without the need for an underlying theory, which becomes more accurate as more terms are integrated into it (\cf \cite{menke1989}, chapters 7.6 to 7.8).
	
\begin{figure}[H]
    \centering
    \includegraphics[width=1\columnwidth]{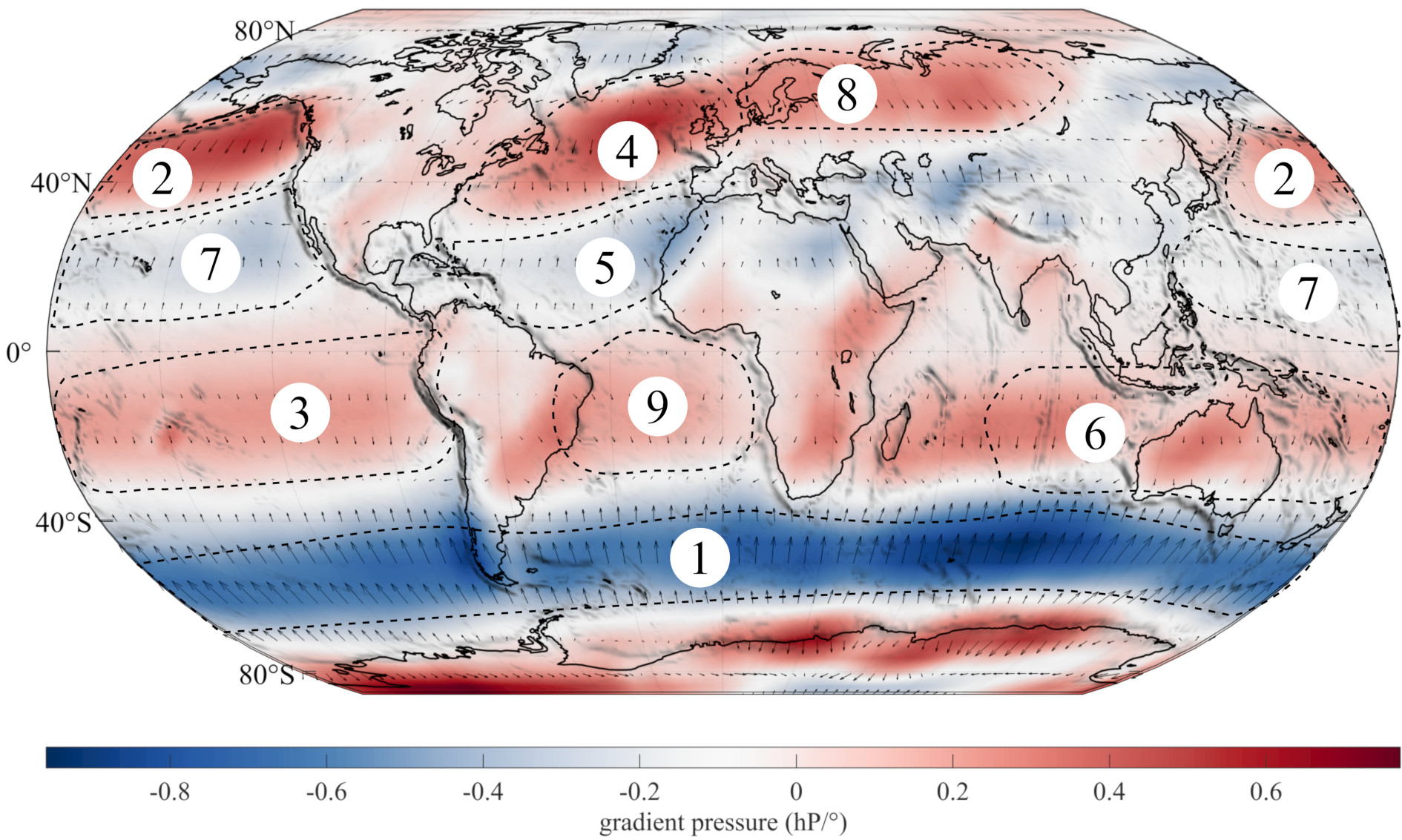}	
    \caption{Map of pressure and topography gradients (Figure \ref{fig:02}), superimposed on the areas where the main climate indices are defined: 1 = South Annular Mode; 2 = Pacific Decadal Oscillation; 3 = El Niño-Southern Oscillation; 4 = North Atlantic Oscillation; 5 = Atlantic Multidecadal Oscillation; 6 = Madden Julian Oscillation; 7 = Pacific North American Index; 8 = Eastern Atlantic/Western Russia; 9 = Tropical Southern Atlantic."}
	\label{fig:03}
\end{figure}

	In Figure \ref{fig:03}, we have outlined the geographical areas where the main climate indices are defined and we have superimposed them on the pressure gradient map (see Figure \ref{fig:02}). The first index, which is the most significant in terms of geographical extent, is the South Annular Mode (SAM, \cf \cite{marshall2003}). The second index relates to the Pacific Decadal Oscillation (PDO, \cf \cite{mantua1997}), followed in the third position by El Niño-Southern Oscillation (ENSO, \cf \cite{philander1983}). In the fourth position, we find the North Atlantic Oscillation (NAO, \cf \cite{jones1997}), followed in the fifth position by the Atlantic Multidecadal Oscillation (AMO, \cf \cite{enfield2001}). In the sixth position, we have the Madden Julian Oscillation (MJO, \cf \cite{zhang2005}), followed by the Pacific North American Index (PNA, \cf \cite{leathers1991}) in the seventh position. In the eighth position, we selected the Eastern Atlantic/Western Russia Index (ER/WR, \cf \cite{krichak2005}), and finally, in the ninth position, we chose to represent the Tropical Southern Atlantic Index (TSA, \cf \cite{enfield1999}).
\newpage

	We could have continued with the forty climate indices that exist today, but that would only have reinforced the main observation made in Figure \ref{fig:03}. As already pointed out, although several parameters are taken into account to construct these climate indices, it is clear that sea-level atmospheric pressure dominates all others. As we will see, this natural division of the indices is not reflected in temperature anomaly maps.

	To our knowledge, we may have been the first to detect and extract the full series of quasi-periodic variations that had already been observed in sunspots (\cf \cite{lemouel2019a}) from the entire set of climate indices. Several studies have since confirmed our findings (\eg \cite{ludecke2020,connolly2021,mares2022,ormaza2022, tacza2022}).

	However, we wish to update these conclusions, with an important remark. All climate indices appear to be closely connected by their geographical distributions (Figure \ref{fig:03}). Pressure anomalies are well defined over a period of observation of just over 170 years. In a series of articles (\cf \cite{lopes2017,lopes2021,lopes2022c}), we have shown, based on \Lagrange 's mechanics (\cite{lagrange1788}) and \Laplace ’s treatise (\cite{laplace1799}), that it was possible to find the numerical values of the periods of revolution of the planets involved in the motion of Earth's rotation pole, due to a simple exchange of angular momentum between these planets and Earth (\cf \cite{milankovic1920,lopes2022d}). Similarly, we have shown that the (famous) 11-year period can be found in several Earth-bound time series such as the length of the day (\eg \cite{lambeck2005}, chapter 5.2; \cite{lemouel2019b}), in global volcanic eruption records from 1700 to the present day (\cf \cite{lemouel2023}), or in the growth rate of tree rings in an entire forest of Tibetan junipers (\cf \cite{courtillot2023b}). This consistently recurring period is present in the evolution of all phenomena that record various aspects of what is commonly referred to as climate (\cf \cite{scafetta2008,scafetta2010,scafetta2012}). This 11-year period, clearly present in sunspots (\cf \cite{schwabe1844,courtillot2021}), is nevertheless absent from temperature curves (\eg \cite{lemouel2020a,courtillot2023a}).

\section{\label{sec07} Rainfall}
\begin{figure}[H]
    \centering
    \includegraphics[width=1\columnwidth]{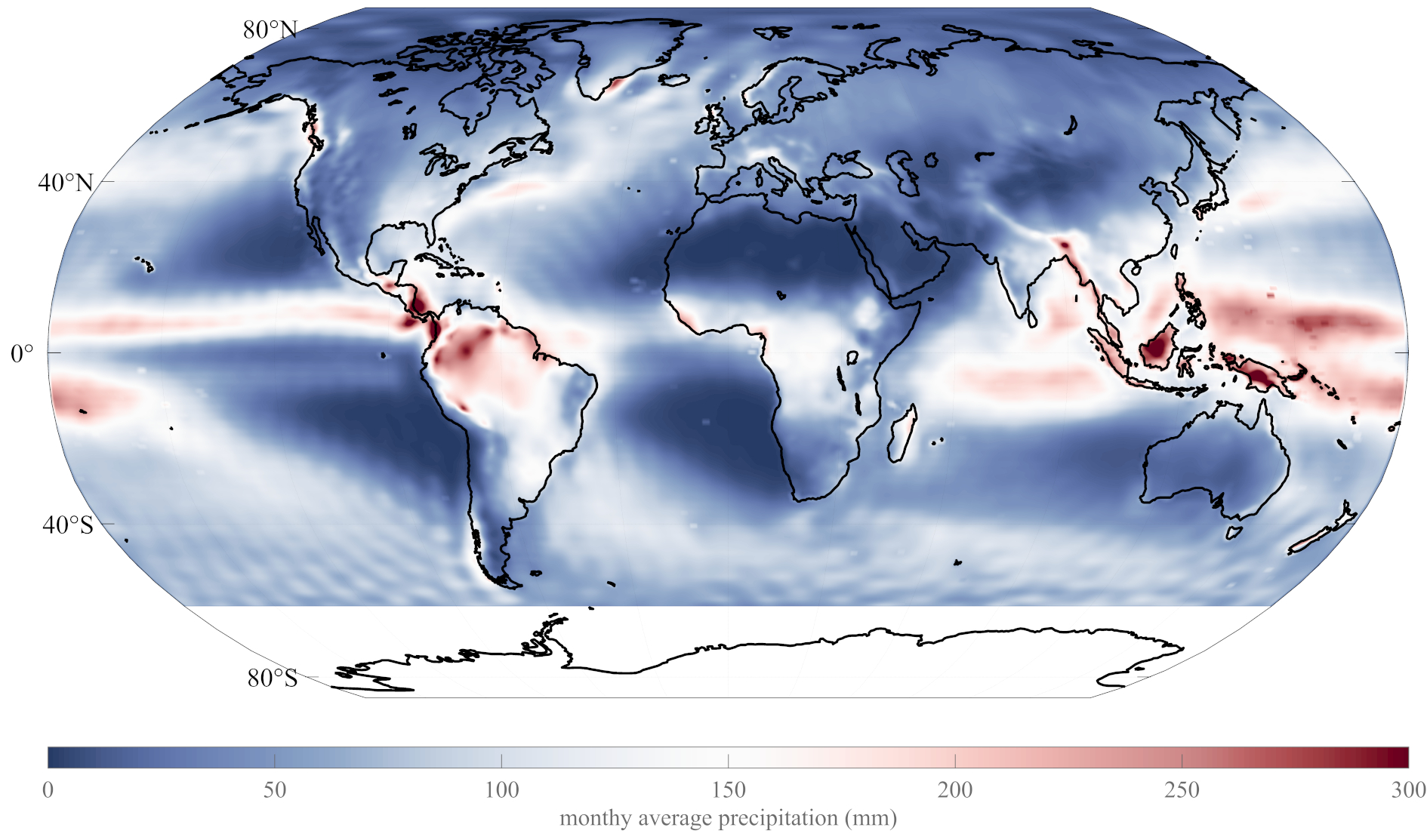}	
    \caption{Mean monthly rainfall map across the globe.}
	\label{fig:04}
\end{figure}

	Another parameter that is part of the list distinguishing meteorology from climate is rainfall. There is a series of data in the form of high-resolution monthly maps (0.5\degree by 0.5\degree, between latitudes 60\degree S and 90\degree N), covering the period from 1948 to 2016. These observations are made available by the Hydrology Laboratory at the University of Southampton\footnote{https://hydrology.soton.ac.uk/data/pgf/}. We present the average monthly rainfall map since 1948 in Figure \ref{fig:04}.	

	The first obvious observation is the presence of a zonal band of high rainfall, located near the equator and almost continuous. Moving from east to west, one can sequentially notice a continuous red band stretching from 180\degree W to Central America, a high rainfall area above the Amazon rain-forest, characterized by a single rainfall climate index, namely the Northeast Brazil Rainfall anomaly (\eg \cite{hastenrath1993}). This band tends to disappear above the Atlantic. A more intense rainfall zone then reappears along the coasts of Liberia and Sierra Leone, extending over the African rain-forest, the southern Sahel, and stretching to about 17\degree S. This zone practically covers the entire extent of the continent's forest. Approaching Ethiopia and Somalia, this anomaly disappears only to reappear in the Indian Ocean south of Asia, clearly present above all the islands of Indonesia. Another interesting observation is that, at least during the period from 1948 to 2016, no notable anomaly appears on average over Asia, especially concerning the Asian monsoon. Thus, the alternation between the wet and dry seasons mitigates the variations on this rainfall map. Broadly speaking, and as expected, rainfall is sensitive to pronounced variations in topography (\eg above Burma, the Himalayas, or the Alps and the Andes; Figure \ref{fig:04}). The driest areas (dark blue) are located on either side of the high rainfall zones (red). The overall structure is not symmetrical about the equator but shifted by about 10\degree to 15\degree, reflecting the distribution of continents.
	
\begin{figure}[H]
    \centering
    \includegraphics[width=1\columnwidth]{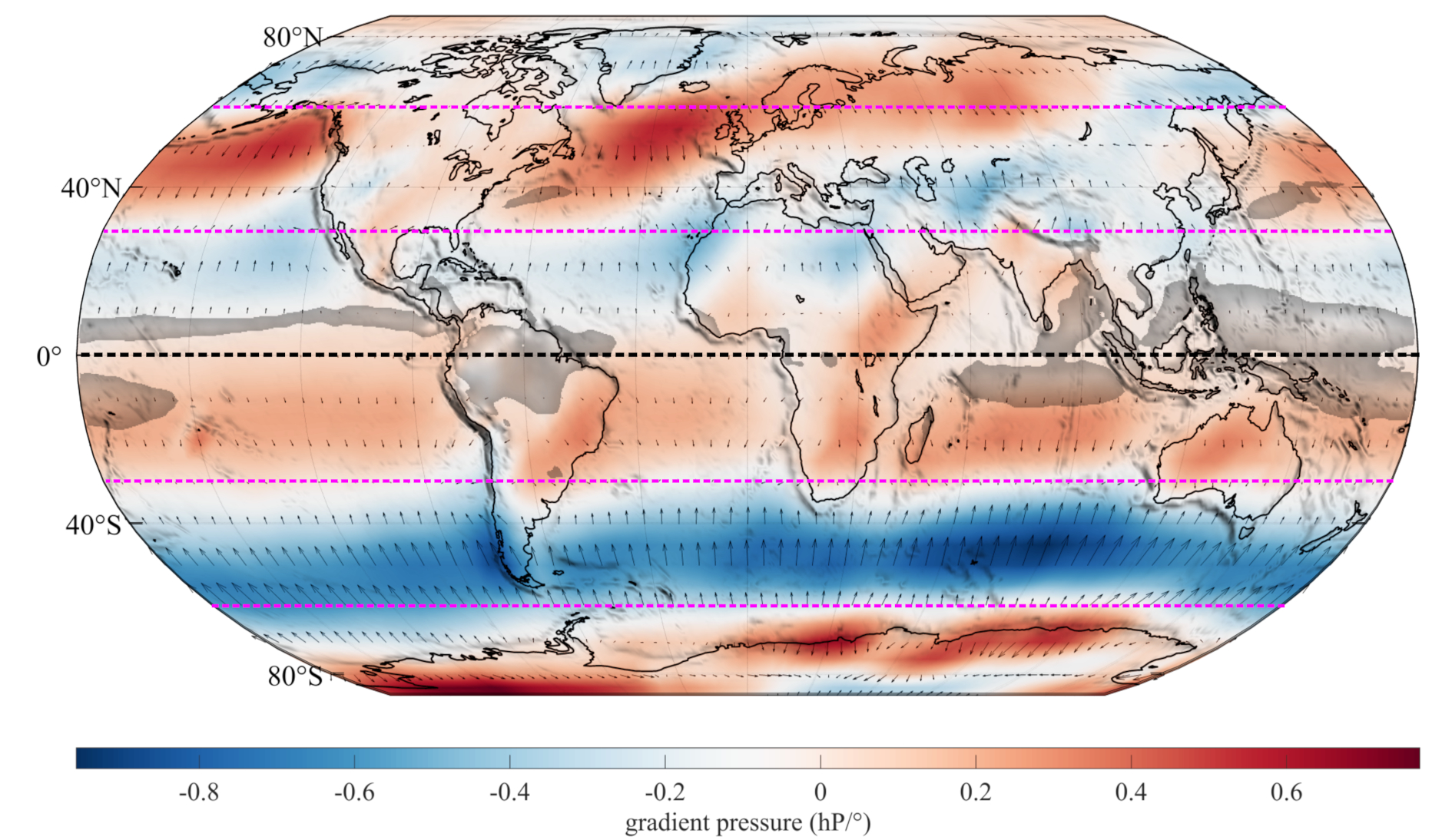}	
    \caption{Overlay on the pressure gradient map (\cf Figure \ref{fig:02}) of precipitation exceeding 150 mm per month, in grayscale.}
	\label{fig:05}
\end{figure}

	The second observation concerns the apparent complementarity between the rainfall map and the pressure gradient map (see Figure \ref{fig:02}). To better appreciate this complementarity, we extracted from Figure \ref{fig:04} the areas where rainfall exceeded 150 mm per month, and we superimposed them (in grayscale) on the pressure gradient map (Figure \ref{fig:05}).
	
	On this Figure \ref{fig:05}, that does not appear to have been discussed in the literature and that we consider to be an original result, we can see that the heavy precipitations, that we have observed to be generally zonal and equatorial, occur in regions where the pressure gradient is nearly zero or close to zero, indicating areas with pressure extrema. Finally, we note that the precipitation band in the Pacific between 180\degree W and Central America is bounded by the ENSO (El Niño Southern Oscillation) and PNA (Pacific-North American) indices, while the precipitation in South Asia and the Pacific is found north of the MJO (Madden-Julian Oscillation) index and is constrained to the west of Indonesia by India and to the west by the PNA.
	
\section{\label{sec08} Temperature Anomaly Maps}
\begin{figure}[H]
    \centering
    \includegraphics[width=1\columnwidth]{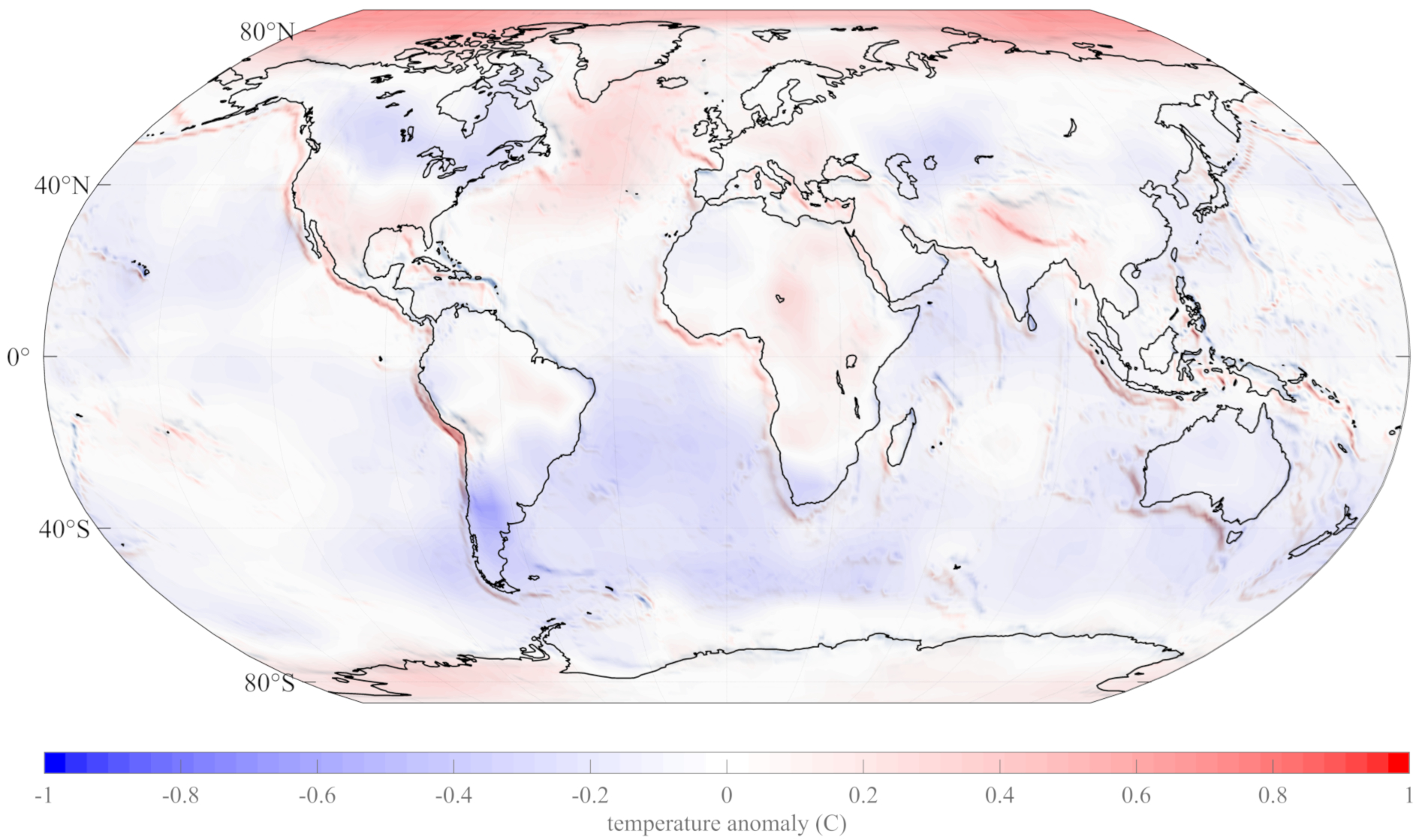}	
    \caption{Map of mean  temperature anomalies since 1850.}
	\label{fig:06}
\end{figure}

	Figure \ref{fig:06} shows the mean temperature anomaly map since 1850 with a spatial resolution of 5\degree by 5\degree. We have overlaid it with topography gradients. The temperature anomaly is negative or null in the oceans (except the North Atlantic), Canada, South America, Central Asia, and Australia. Conversely, it is positive along the ocean-continent boundary from Alaska to Chile, in Central Africa, the Himalayas, and at high latitudes, especially in the Arctic region. Like the previous figure, it appears that the distribution of temperature anomaly patterns does not follow a pattern similar to those of winds, pressures, pressure gradients, or precipitation. For example, the presence of a large temperature anomaly dipole, with a positive zone over the United States and a negative zone over Canada, does not correspond to anything in our other maps. Similarly, the Eastern Atlantic/Western Russia climate index (\cf Figure \ref{fig:03}, number 08) shows no correlation with the temperature anomaly (\cf Figure \ref{fig:06}). 
	
	While data related to temperature anomalies are regularly discussed and even criticized (\eg \cite{soon2023}), it's important to keep in mind that they do not constitute direct measurements. These maps are constructed from actual measurements but undergo various filtering and homogenization processes. For more detailed information regarding these different treatments, especially concerning the HadCrut data in Figure \ref{fig:06}, we refer the reader to the works of \Rayner (\cf \cite{rayner1996, rayner2003,brohan2006,osborn2014,osborn2021}).
	
\section{\label{sec09} Data and Measurements. Variability of HadCrut Trends}
	The meaning of the terms 'data' and 'measurement' needs to be clearly distinguished. Data originate from a sensor, typically a physical one, and are associated with a margin of error. A measurement is the mathematical distance that separates a piece of data from a theoretical model or equation. An observation is simply that, a simple observation. It is understood that, regardless of technological advancements in data acquisition or observational instruments, what was estimated in the past retains its value. On the other hand, a measurement depends on a theory or a model. If these evolve over time, for whichever reason, it is obvious that the measurements, that are previously calculated quantities, will need to be recalculated and thus modified.
\newpage	
\begin{figure}[H]
    \centering
    \includegraphics[width=1\columnwidth]{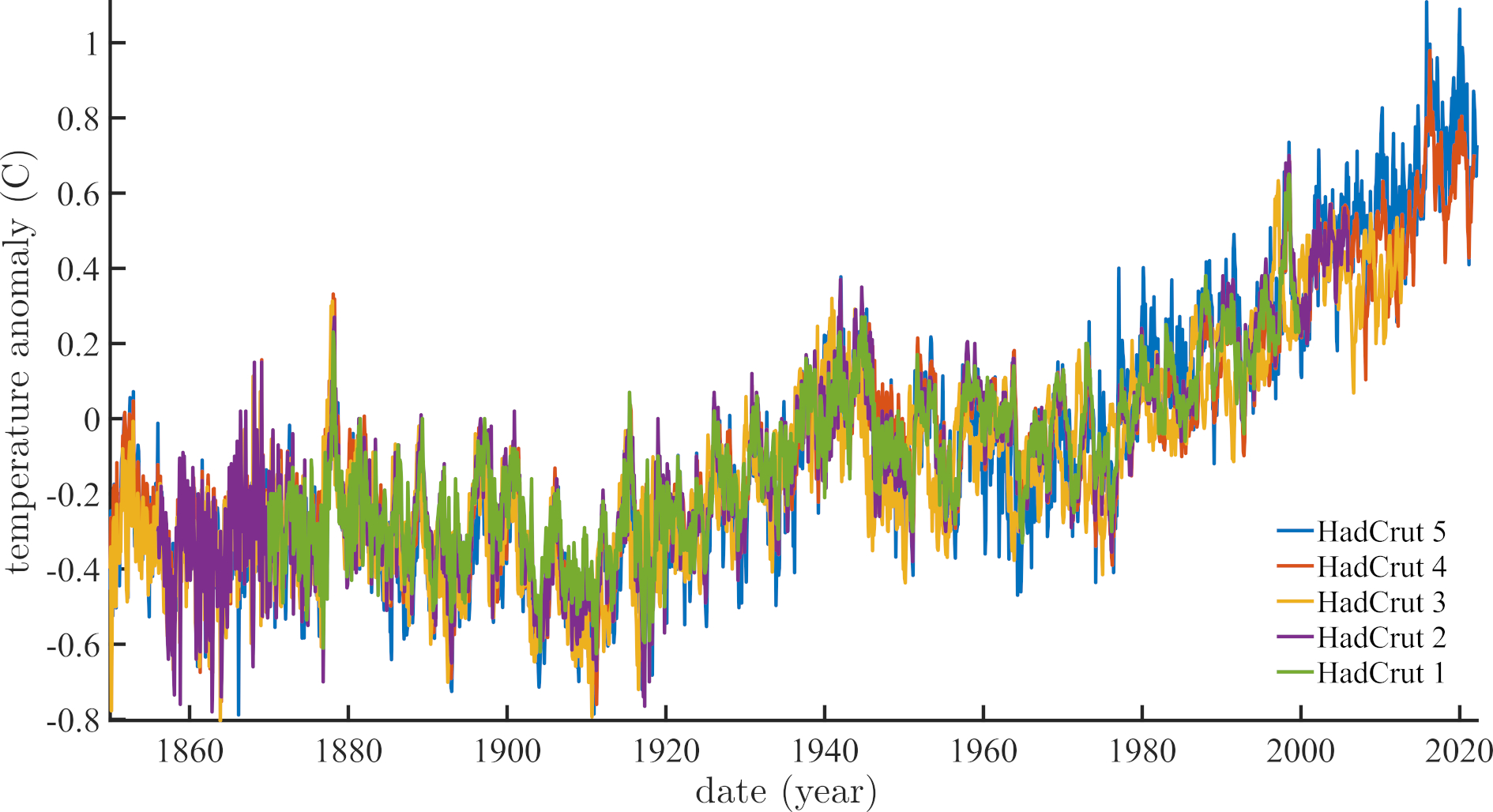}	
    \caption{Overlay of the 5 sets of HadCrut data}
	\label{fig:07a}
\end{figure}	
\begin{figure}[H]
    \centering
    \includegraphics[width=1\columnwidth]{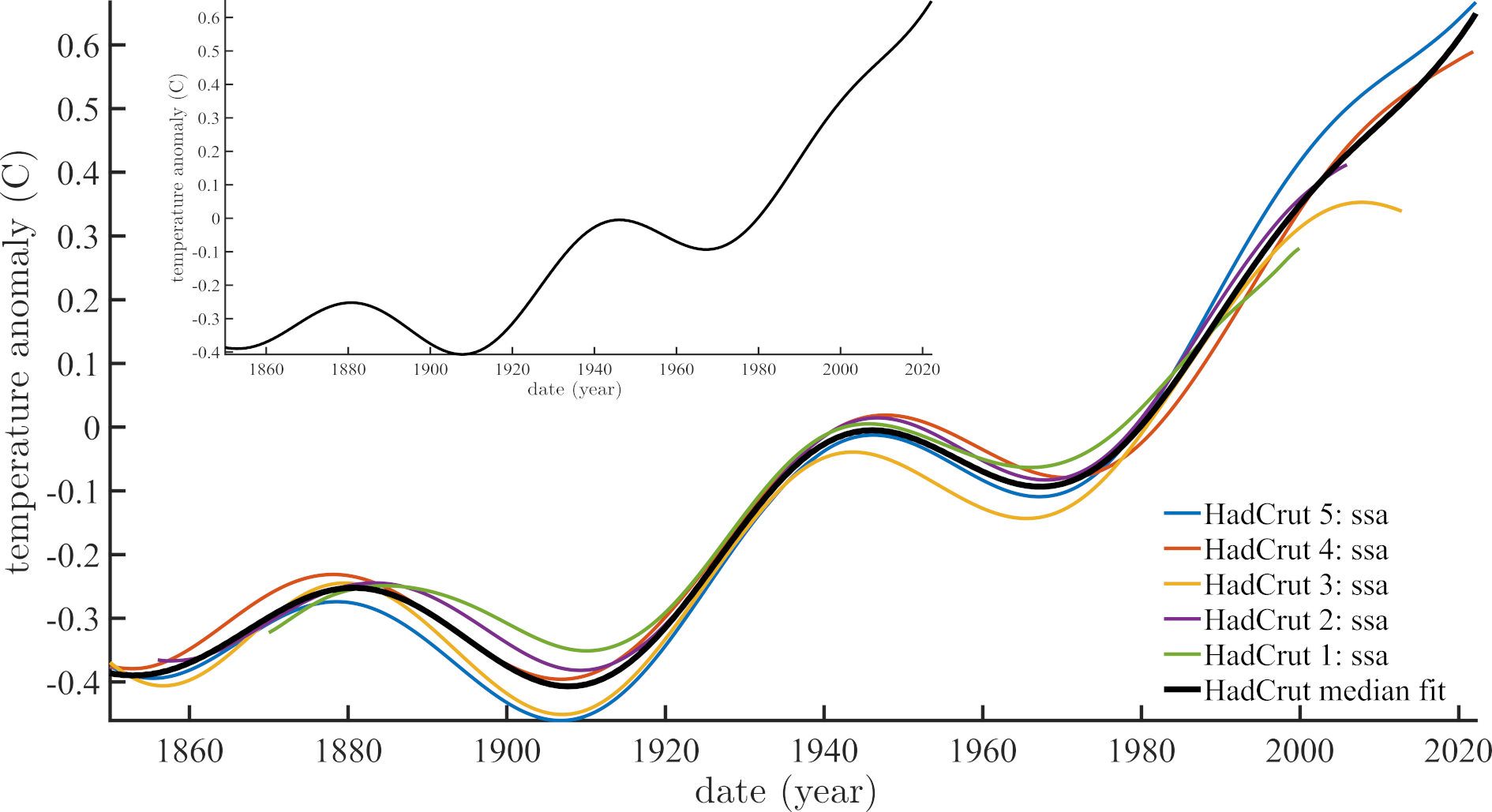}	
    \caption{Overlay of the 5 trends determined by SSA (from \cite{lopes2022a}).}
	\label{fig:07b}
\end{figure}	
	
	This is illustrated by Figure \ref{fig:07a} (from \cite{lopes2022a}, Figure 01), where the 5 HadCrut datasets are overlaid on the same graph. These datasets cover the following time periods, in chronological order: 1870-2000 (HadCrutv, \cite{rayner1996}), 1856-2006 (HadCrut2, \cite{rayner2003}), 1850-2014 (HadCrut3, \cite{brohan2006}), 1850-2021 (HadCrut4, \cite{osborn2014}), and finally 1850-2023 (HadCrut5, \cite{osborn2021}). The different sets of measurements do not strictly overlap in the past; however, they remain fairly close. Nevertheless, these differences can have significant implications for forecasting. We have addressed this issue, both for trends and annual oscillations (in \cite{lopes2022a}), and revisit now some key points. If one examines closely the curves for HadCrut2 (purple), HadCrut3 (yellow), and HadCrut4 (red), the end of each time series exhibits a plateau at different dates and durations (Figure 07b). This plateau has led to discussions of what has come to be known as the "climate hiatus" (\eg \cite{karl2015,lewandowsky2015}). In particular, the HadCrut3 curve (yellow) shows a stable plateau from 1990 to 2010. Many professional observers might have been tempted to conclude that the temperature anomaly curve had stopped growing during that time. However, one year later, with the introduction of the HadCrut4 curve (red), this plateau had disappeared (Figure \ref{fig:07b}).	
	
\section{\label{sec10} Temperature Anomalies as a Function of Latitude}
\begin{figure}[H]
    \centering
    \includegraphics[width=1\columnwidth]{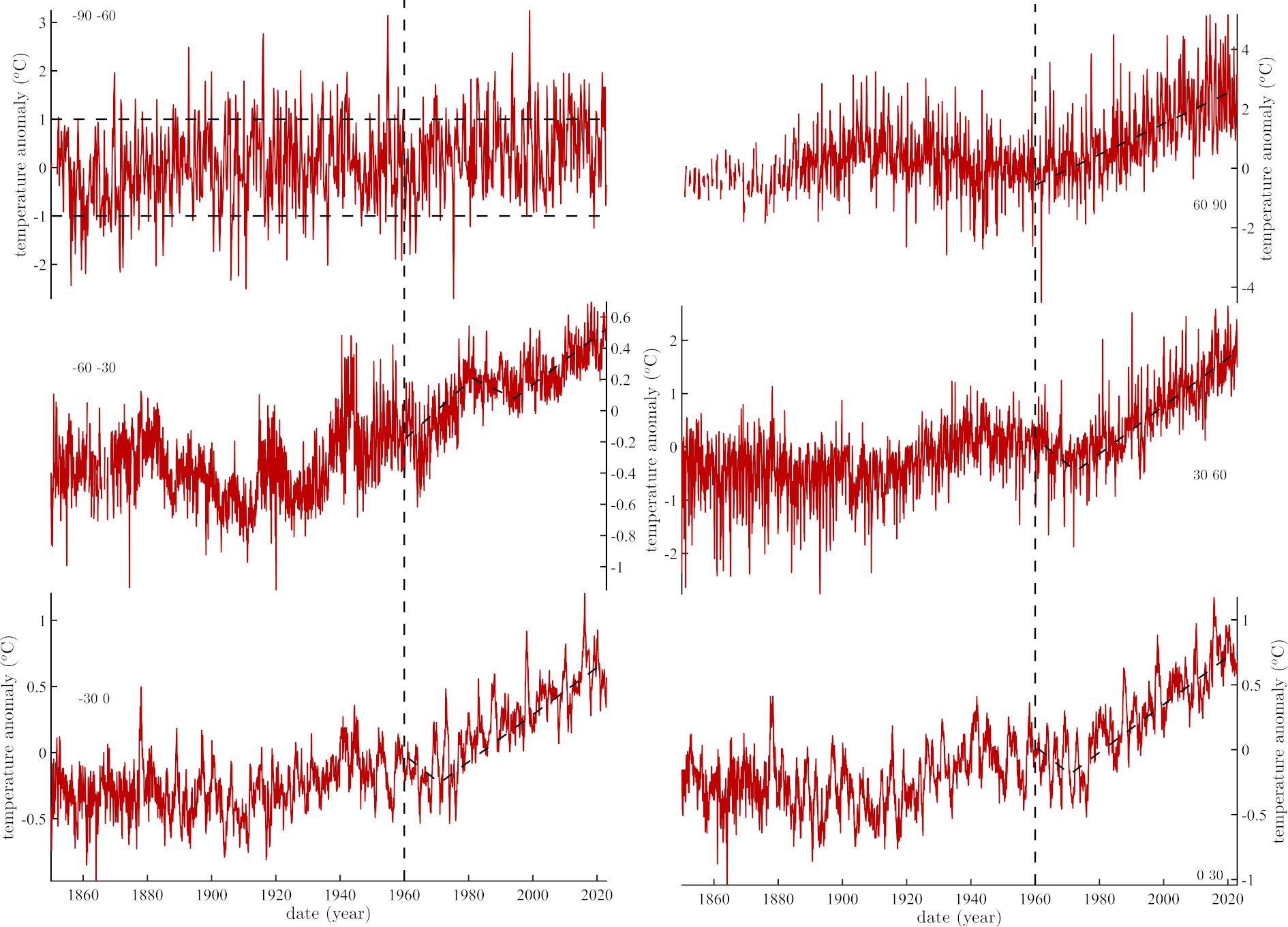}	
    \caption{Temperature anomaly evolution curves by latitude band. At the top, from left to right: between 90°S and 60°S, between 90°N and 60°N. In the middle, from left to right: between 60\degree S and 30\degree S, between 60\degree N and 30\degree N. At the bottom, from left to right: between 30\degree S and 0\degree, between 30\degree N and 0\degree. Vertical dashed lines indicate a time reference (1960)}
	\label{fig:08}
\end{figure}

	The mean temperature anomaly curve (\cf Figure \ref{fig:07a}), like all average curves, conceals subtleties and disparities. Similar to what we have done for previous climate variables, we divided the anomaly map into six latitude bands of 30\degree width each. For each of these bands, we assessed the evolution of the average temperature over time using the most recent HadCrut series, version 5 (blue in Figure \ref{fig:07b}). The temperature anomaly curves for each latitude band are shown in Figure \ref{fig:08}.

	A careful examination of the HadCrut (v5) average data, broken down into six latitude bands, leads to two important observations. First, there are significant differences between the two hemispheres. In the polar Southern Hemisphere (top-left curve), the temperature anomaly has remained relatively stable since 1850, staying within a range of $\pm$1\degree C. However, in the polar Northern Hemisphere (top-right curve), there is a half-oscillation with an amplitude of +2\degree C before 1960, followed by a modest linear increase of approximately +2\degree C\footnote{The following description is a preliminary draft and is not the only possible one (see Courtillot et al. \cite{courtillot2013} and Le Mouël et al. \cite{lemouel2020a}). It helps to establish the order of magnitude of temperature differences and their durations}. However, over the past thirty years, the signal-to-noise ratio has shifted in favor of noise, to the point when 'peaks' exceeding +4\degree C appear quite regularly. These peaks can be observed globally in the entire Northern Hemisphere, and they are present on all three curves in the right column at the same moments. This latter point is somewhat surprising because these ‘peaks’ are generally attributed to more or less significant variations associated with El Niño (\eg \cite{masson2021}, Chapter 2.4, page 370, and Paragraph 2.4.2, page 371, regarding El Niño). The fact that temperature anomalies have remained relatively constant above Antarctica since 1850, while they have increased by 2 $\pm$ 2\degree C above the Arctic since 1970, could explain why sea ice is melting in the Northern Hemisphere while forming in the Southern Hemisphere (\cite{cavalieri2012,parkinson2012,parkinson2019,lemouel2021,lopes2023a}).
\newpage	

	In the 30\degree (N/S) - 60\degree (N/S) latitude bands (see the middle curves in Figure \ref{fig:08}, the behaviors are still very different. In the Southern Hemisphere (left curve), one could estimate that one has emerged from "natural variability" around the 1990s. The subsequent growth is approximately linear. Between 1990 and the Present, the temperature anomaly would have risen from +0.1\degree C to +0.5\degree C, an increase of +0.4\degree C. During the same period, in the Northern Hemisphere, the temperature anomaly would have risen from about 0\degree C to +1.5\degree C, which is three times larger.
	
	This contrast between the two hemispheres raises questions about the physical meaning of the notion of mean temperature and the 'average' greenhouse effect. If we crudely average the variations in temperature anomalies since 1990, we observe that in the Southern Hemisphere, between 90\degree S and 30\degree S, the anomaly has increased on average by about 0.2\degree C ( (0\degree C + 0.4\degree C)/2 ), whereas in the Northern Hemisphere, it has increased by about 2\degree C ( (2\degree C + 2\degree C)/2 ), which is ten times larger. 'Climate change' primarily affects the Northern Hemisphere ( $ > $ 60\degree N).
	
	The values around the equator, shown in the bottom curves of Figure \ref{fig:08}, are very similar. Since 1990, the temperature anomaly has increased by only 0.5\degree C.
	
\section{\label{sec11} Temperature Anomalies as Functions of Time}
	In seeking to provide a physical interpretation for the temperature mean trends, one can propose the following: the Earth's lower atmosphere is gradually warming from the North Pole to the South Pole. In Figure 9, we have depicted the evolution of the map of average temperature anomalies in 20-year intervals since 1850. The white areas, prominently visible in the first two maps at the top left, as well as above Antarctica in the first five maps, indicate the absence of measurements. It is worth noting that the number of longitude-latitude pairs for which we have temperature measurements only covered about 50\% of the Earth's surface in 1850, decreased to 40\% after 1860, then increased in steps to reach around 70\% by 1960. It was only with the advent of satellites that the coverage finally reached 100\% of the Earth's surface (\eg \cite{courtillot2023a}, Figure 01, top curve).
\begin{figure}[H]
    \centering
    \includegraphics[width=1\columnwidth]{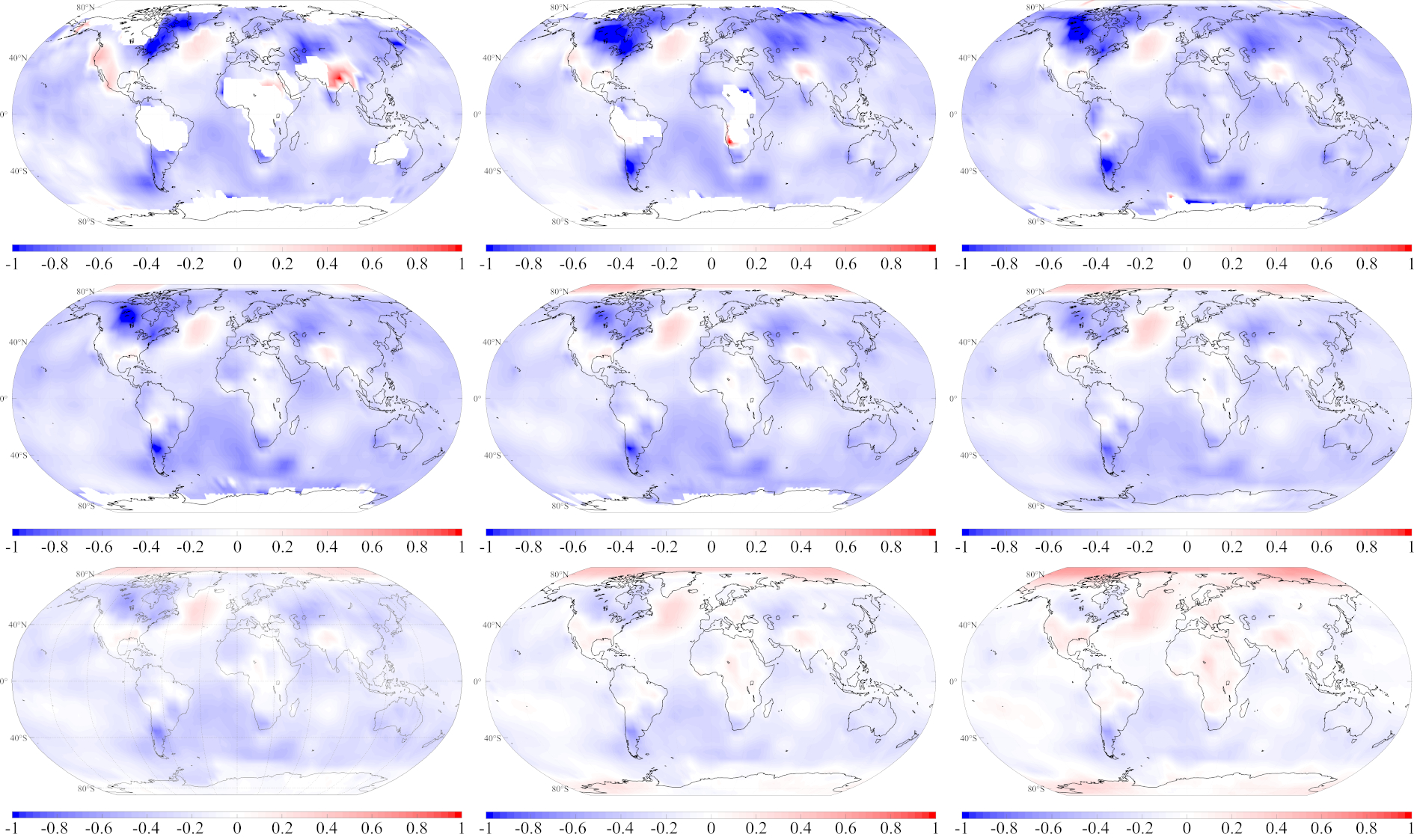}	
    \caption{Evolution of the median map of temperature anomalies. Top row, from left to right: between 1850 and 1870, between 1850 and 1890, between 1850 and 1910. Middle row, from left to right: between 1850 and 1930, between 1850 and 1950, between 1850 and 1970. Bottom row, from left to right: between 1850 and 1990, between 1850 and 2010, between 1850 and 2023. The color code remains the same for all 9 maps, linearly ranging from -1\degree C (blue) to +1\degree C (red).}
	\label{fig:09}
\end{figure}	

	In this Figure \ref{fig:09}, we have retained the color scale, ranging from $\pm$ 1\degree C, which is consistent with the temperature range in the 30\degree S-90\degree  S band. It is clear, when looking at these nine maps, that indeed the Earth is warming, and it has been doing so southward from the Arctic.

\section{\label{sec12} Discussion}
	Modern definitions of climate and meteorology were established during two international meteorological conferences. Climatologists and meteorologists agreed on a separation of 30 years, corresponding to the BEL cycle (\cite{bruckner1890}). They did not take into account sea levels in their definition, but things have changed significantly (\eg \cite{lamb2013} or \cite{masson2021}, Chapter 3, page 426). \Douglas (\cite{douglas1997}) begins his article with the following sentence: 'It is well established that sea level trends obtained from tide gauge records shorter than about 50-60 years are corrupted by interdecadal sea level variation'. There appears to be a contradiction between the duration of the BEL cycle and the one mentioned by \Douglas (\cite{douglas1997}), which is twice as long. Therefore, it is possible that conducting climate studies over a slightly longer period than 30 years and attempting to estimate sea level since 1992 using satellites may lead to confusion, potentially conflating a 60-year cycle with exponential growth (\cf \cite{wagenaar1975}). 

	The maps of mean sea-level pressures, their gradients, winds, and precipitation (\cf Figures \ref{fig:01} to \ref{fig:05}) show that these phenomena exhibit remarkably stable patterns, both in time and space. Furthermore, these patterns do not seem to inherently require the consideration of temperature variations to explain their nature. This is a well-established but unfortunately often overlooked fact in modern times, as we clearly illustrate on our maps: sea-level pressure variations, their gradients, as well as what climatologists call climate indices (\eg \cite{lemouel2019a}), are primarily determined by the gradients of terrestrial topography (\cf Figures \ref{fig:03} and \ref{fig:05}) and are thus connected by the same general physics. One can follow and verify \Laplace 's reasoning (\cite{laplace1799}). One can also recall the words of \Fourier (\cite{fourier1827}, page 108): 'The intensity and distribution of heat on the surface of these bodies result from the distance to the Sun, the inclination of the axis of rotation to the orbit, and the state of the surface'.
	
	The patterns we have highlighted, which are remarkably stable, are not found in the map of surface temperatures (\cf \cite{lembrechts2022}), nor even in those of temperature anomalies (\cf Figures \ref{fig:06} and \ref{fig:09}). It appears that temperature and its evolution are largely unrelated to other physical aspects of the climate.
	
	Two paradigms regarding the nature of temperature on Earth have long been in opposition. The variations over time in the temperature at the Earth’s surface can be attributed either to the \Milankovic (\cite{milankovic1920}) cycles or to the greenhouse effect (\cite{fourier1827,arrhenius1896}). It is accepted that significant temperature and climate variations observed during glacial cycles ($\sim$ 20,000 years, 400,000 years, etc.) are linked to variations in the obliquity, precession, and eccentricity of the Earth's axis of rotation in our solar system (\eg \cite{agassiz1837,adhemar1860,croll1864,milankovic1920}, as detailed in \cite{lopes2021b}). For shorter periods, less than a hundred years, the greenhouse effect is invoked, particularly variations in the  concentration in the atmosphere. Regarding this, one can read in \Arrhenius (\cite{arrhenius1896}), page 270: '\textit{The world's present production of coal reaches in round numbers 500 millions of tons per annum, or 1 ton per km$^{2}$ of the earth's surface. Transformed into carbonic acid, this quantity would correspond to about a thousandth part of the carbonic acid in the atmosphere}'.
	
\begin{figure}[H]
    \centering
    \includegraphics[width=1\columnwidth]{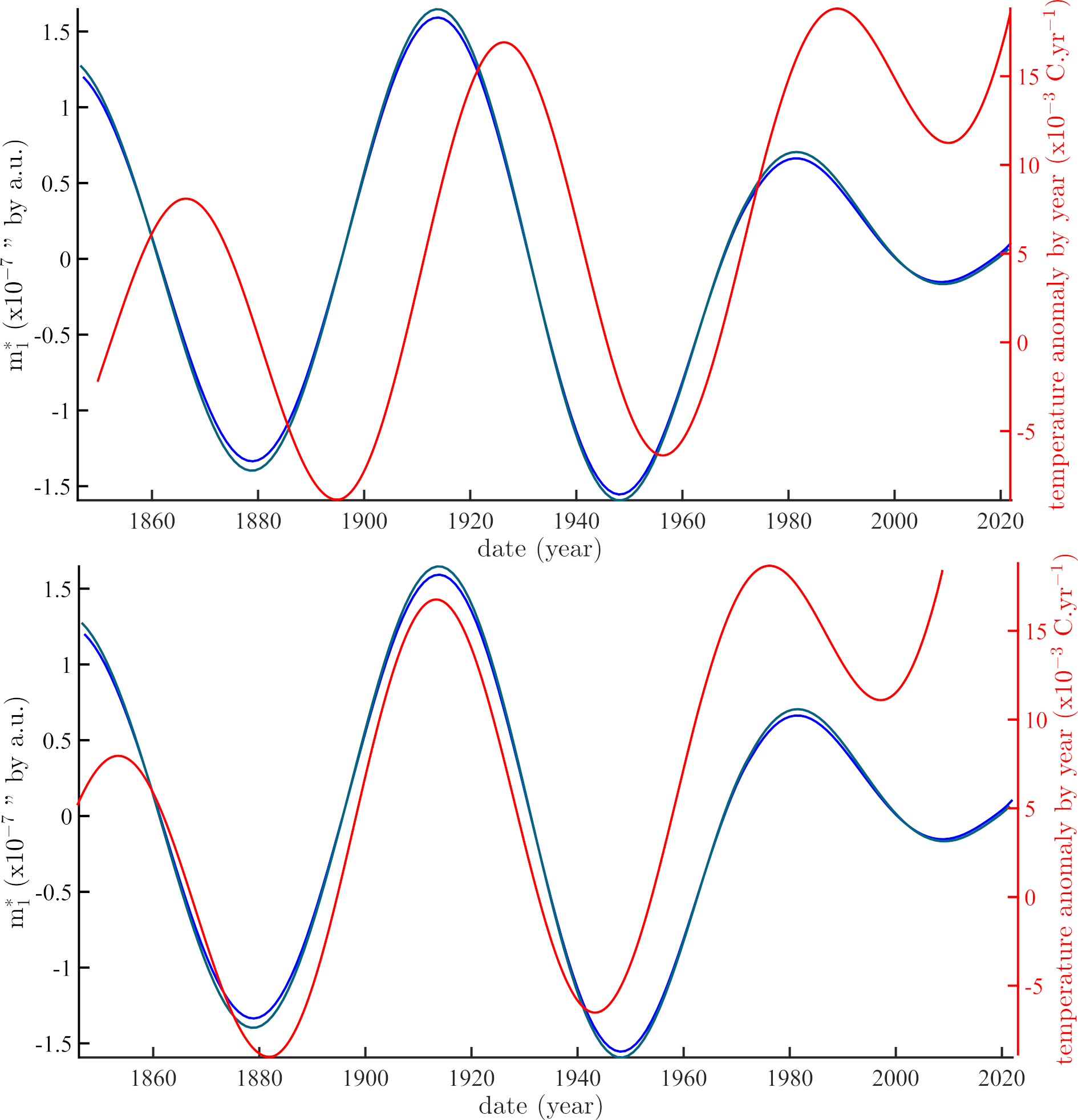}	
    \caption{At the top (in blue and green), the inverse square of the variations in the distance from the Earth's rotation pole to the Sun during the Summer and Winter solstices. In red, the time derivative of the temperature anomaly trend extracted from HadCrut5. At the bottom, the same curves corrected for a phase shift (15 years = 60 years/4). Figure extracted from \cite{lopes2022a}, Figure \ref{fig:10}.}
	\label{fig:10}
\end{figure}

	It is surprising to note that when it comes to studying temperature variations over less than two centuries, the Milanković theory is rarely, if ever, invoked. This means that, in the perspective of modern climatologists, one transitions instantly from a warm period to an ice age, and then to another warm period. In \Lopes \cite{lopes2022a}, we extended \Milankovic (\cite{milankovic1920})'s theory to the past 250 years. \Milankovic 's general relationship (\cite{milankovic1920}, page 15, equation 20) relates the temporal variation (derivative) of the amount of insolation received at the Earth's surface to the variation in the inverse of the Earth-Sun distance squared. This relationship leads to a perfect correspondence (in the sense of \Keogh \cite{keogh2002,lin2003,keogh2005}) between the HadCrut5 trend and the variations in the Earth's distance since 1850 (see Figure \ref{fig:10}).
	
	In this paradigm, it seems that \Laplace (\cite{laplace1799}) was the first to address the question of the Sun's action on Earth. The Sun is constant, at least for the past 200 years during which we have measured its radiation. This radiation puts the molecules of Earth’s atmosphere in motion through simple thermodynamic effects. However, the issue is that, as highlighted by \Laplace (\cite{laplace1799}) (as well as \cite{legendre1785,lagrange1788,poisson1826,fourier1827}), the cause of the rotations of the solid Earth, oceans, and atmosphere is mainly found in celestial mechanics. Thus, whether solar radiation is constant or not, exchanges of angular momentum are constantly occurring between the planets themselves (\eg \cite{lopes2022d}), between the oceans and our solid Earth (\Laplace \cite{laplace1799}, shows that oceans behave like solids in rotation at the first order), and between the atmosphere and the solid Earth. However, the mass of Earth’s atmosphere ($5.29 * 10^{18}$ kg) is negligible compared to that of the oceans ($1.35 * 10^{21}$ kg) and even more so compared to that of the entire planet ($5.98 * 10^{24}$ kg).
	
	In these conditions, as the atmosphere cannot exchange (strong enough) momentum with the solid Earth, its effect becomes insensitive to the rotation of the latter, and thus the molecules in motion due to solar radiation compensate for their velocities: 'We are therefore assured that as the analyzed winds decrease this motion, the other atmospheric movements occurring beyond the tropics accelerate it by the same amount'. Thus, we lose the spatiotemporal correlation between ground temperature or its variations (independently of the cause) and other climate physics. This is precisely what all the maps we have presented teach us. Therefore, this minimizes the influence of the atmosphere and its physico-chemical properties, as illustrated by the success of Milanković's mathematical theory of climate, a theory in which there is no atmosphere.

	This raises questions about the importance and significance of the greenhouse effect and albedo. Both of these effects appear to be of secondary importance in our view. To support this idea, which is already evident in the data (see all our maps), we begin with Arrhenius's seminal article from 1896. In this article, the author hopes that his theory, which is based on the effect of  on temperature, can explain ice ages and paleo-climates. He compares his hypothesis to those existing at his time, drawing on a synthesis article by de \Marchi (\cite{marchi1895}), in which various topics are addressed, including (on page 273) changes in the position of the equinoxes, variations in the obliquity of Earth's axis of rotation, shifts in the pole positions, and \Milankovic 's mathematical theory of climates. \Arrhenius (\cite{arrhenius1896}) refutes all of de \Marchi 's (\cite{marchi1895}) hypotheses, including \Milankovic ' (\cite{milankovic1920}).

	\Arrhenius 's theory is based on a generalization of the blackbody radiation law, known as Stefan's law (see page 255). He attempts to extend this law to account for the complexity of the Earth's atmosphere, resulting in a complex equation that relates various parameters such as albedo and surface temperature (equation 03, page 256). To develop this equation, the author had to use a large number of constants related to various components of the Earth, biological, chemical, and physical reactions, and so on. Today, we know that most of these "constants" do not exist. Therefore, \Arrhenius 's conclusions from 1896 can no longer be accepted as they are. Furthermore, Arrhenius himself attenuates his conclusions from page 254 onward, including in the synthesis by de Marchi. For example, he writes (pages 254-255): '\textit{This reasoning holds good if the part of the earth's surface considered does not alter its albedo as a consequence of the altered temperature. In that case entirely different circumstances enter. If, for instance, an element of the surface which is not now snow-covered, in consequence of falling temperature becomes clothed with snow, we must in the last formula not only alter   but also ’. As the author himself acknowledges, his reasoning is valid only if ‘the earth's surface considered does not alter its albedo}'. However, volcanoes, the biosphere, soil alterations can all modify the albedo.	
	
	The findings presented in this slightly ‘impressionist’ note confirm the remarkable theoretical works of the \Laplace (\cite{laplace1799,legendre1785,lagrange1788,poisson1826}). Firstly, temperatures are the only variables that do not follow the patterns of the main climate parameters over time and space (see Figures \ref{fig:05} to \ref{fig:06}). Secondly, Earth has been warming over the past century from the North Pole towards the equator (see Figures \ref{fig:09} and \ref{fig:10}). The Earth's axis of rotation has been tilting since 1900 (\eg \cite{lopes2022c}), dragging the magnetic field1 with it (see \cite{lemouel2023b}), triggering volcanic eruptions (see \cite{lemouel2023}), influencing forest life rhythms (see \cite{courtillot2023b}), and even driving sea level changes (\eg. \cite{courtillot2022b}).

 Finally, could it turn out that temperature is neither a forcing factor of climate nor a consequence of the main mechanisms that govern it ? Isn’t \Milankovic (\cite{milankovic1920}) closer to the truth than \Arrhenius (\cite{arrhenius1896}) ?
 
 \newpage

\bibliographystyle{ieeetr}
\bibliography{biblio}
\end{document}